\RequirePackage{luatex85}

\documentclass{article}

\usepackage{arXiv}

\usepackage[utf8]{inputenc} 
\usepackage[T1]{fontenc}    
\usepackage{hyperref}       
\usepackage{url}            
\usepackage{booktabs}       
\usepackage{amsfonts}       
\usepackage{nicefrac}       
\usepackage{microtype}      
\usepackage{lipsum}		
\usepackage{graphicx}
\usepackage{natbib}
\usepackage{doi}

\title{Rayleigh--Bloch waves above the cut-off}

\author{Luke G.~Bennetts \\
	University of Adelaide, Australia \\
	\texttt{luke.bennetts@adelaide.edu.au} 
		\And
	Malte~A.~Peter \\ 
	University of Augsburg, Germany \\
	\texttt{malte.peter@math.uni-augsburg.de}
}

\date{}

\renewcommand{\shorttitle}{Rayleigh--Bloch waves above the cut-off}

\hypersetup{
pdftitle={Rayleigh--Bloch waves above the cut-off},
pdfauthor={Luke G.~Bennetts and Malte A.~Peter},
pdfkeywords={Wave scattering, Rayleigh-Bloch waves, Diffraction grating},
}

\def\shownewold{SHOWING NEW/OLD/DELETED TEXT}

\usepackage{amsmath}

\usepackage{stmaryrd}   
\newcommand{\arrowscross}{\scalebox{0.5}{$\nearrow\hspace{-10.5pt}\searrow$}}

\newcommand{\ie}{i.e.\ }

\renewcommand{\exp}{{\rm exp}}

\newcommand{\wrt}{\,\mathrm{d}}   
    
\newcommand{\ci}{\mathrm{i}}   
\newcommand{\e}{\mathrm{e}}

\usepackage{enumerate}

\usepackage{overpic,rotating}

\newcommand{\beq}{\begin{equation}}
\newcommand{\eeq}{\end{equation}}
\newcommand{\bea}{\begin{eqnarray}}
\newcommand{\eea}{\end{eqnarray}}
\newcommand{\bean}{\begin{eqnarray*}}
\newcommand{\eean}{\end{eqnarray*}}

\newlength{\mylinelength}
\newlength{\mydashlength}
\newlength{\mydashspace}
\newlength{\mychainlengthA}
\newlength{\mychainlengthB}
\newlength{\mychainspace}
\newlength{\mylinethickness}
\newcommand{\mathand}{\quad\textnormal{and}\quad}
\newcommand{\mathandR}{\textnormal{and}\quad}

\newcommand{\mathXR}[1]{\textnormal{#1}\quad}

\usepackage{pifont}
\ifdefined\showmylabels
 \usepackage[inline]{showlabels}
 
\fi

\usepackage[textwidth=1.5cm,textsize=tiny]{todonotes}
\usepackage[normalem]{ulem}
\usepackage{soul}
\ifdefined\shownewold
   \newcommand{\del}[1]{{\color{red}\sout{#1}}}
\else
   \newcommand{\del}[1]{\ignorespaces}
\fi
\ifdefined\shownewold
   \newcommand{\deleqn}[1]{\\\del{\parbox{\textwidth}{#1}}}
\else
   \newcommand{\deleqn}[1]{\ignorespaces}
\fi
\ifdefined\shownewold
   \newcommand{\new}[1]{{\color[rgb]{0,0,1}#1}}
\else
   \newcommand{\new}[1]{#1}
\fi
\ifdefined\shownewold
   \newcommand{\old}[1]{{\color[rgb]{0.5,0.5,0.5}#1}}
\else
   \newcommand{\old}[1]{\ignorespaces}
\fi
\ifdefined\shownewold
   \newcommand{\delref}[1]{\del{\parbox{\figwidth\columnwidth}{#1}}}
\else
   \newcommand{\delref}[1]{\vspace{-12pt}}
\fi
\ifdefined\shownewold
   \newcommand{\lu}[1]{{\color[rgb]{1,0,1} #1}}
   \newcommand{\luchange}[2]{\del{#1}\new{#2}}
   \newcommand{\luleft}[1]{{\color[rgb]{1,0,1}$\longleftarrow$#1}}
   \newcommand{\luright}[1]{{\color[rgb]{1,0,1}#1$\longrightarrow$}}
\else
   \newcommand{\lu}[1]{\ignorespaces}
   \newcommand{\luchange}[1]{\ignorespaces}
   \newcommand{\luleft}[1]{\ignorespaces}
   \newcommand{\luright}[1]{\ignorespaces}
\fi
\usepackage{tikz,pgfplots}
\usetikzlibrary{3d,calc,patterns,decorations.pathmorphing,decorations.markings,decorations.pathmorphing,decorations.text,angles,quotes,arrows,arrows.meta,backgrounds,positioning,shapes,shapes.geometric,intersections}
\tikzset{snake it/.style={decorate, decoration=snake}}

\newlength{\figurewidth}
\newlength{\figwidth}
\newlength{\colsep}
\newlength{\figureheight}
\newlength{\myxunit}
\newlength{\rad}

\usepackage{xcolor}
\definecolor{darkgreen}{rgb}{0,0.55,0}
\definecolor{midgreen}{rgb}{0,0.8,0.2}
\definecolor{magenta}{rgb}{1,0,1}
\definecolor{purple}{rgb}{0.5,0,0.5}
\definecolor{darkorange}{rgb}{1,0.55,0}
\definecolor{maroon}{rgb}{0.5,0,0}
\definecolor{olive}{rgb}{0.5,0.5,0}
\definecolor{midgrey}{rgb}{0.5,0.5,0.5}
\definecolor{lightgrey}{rgb}{0.75,0.75,0.75}
\definecolor{matlabblue}{rgb}{0,0.447,0.741}
\definecolor{matlabred}{rgb}{0.85,0.325,0.098}
\definecolor{lightblue}{rgb}{0,0.5,1}
\definecolor{darkgrey}{rgb}{0.25,0.25,0.25}
\definecolor{teal}{rgb}{0,0.5,0.5}
\definecolor{navy}{rgb}{0,0,0.5}
\definecolor{goldenrod}{rgb}{0.85,0.6,0.1}

\begin{document}
\maketitle

\begin{abstract}
	Extensions of Rayleigh--Bloch waves above the cut-off frequency are studied via the discrete spectrum of a transfer operator for a generalised channel containing a single cylinder.
Their wavenumbers are shown to become complex-valued and an additional pair of wavenumbers to appear.
For small to intermediate radius values, the extended Rayleigh--Bloch waves are shown connect the Neumann and Dirichlet trapped modes, then embed in the continuous spectrum. 
Rayleigh--Bloch waves vanish as frequency increases but reappear at high frequencies for small and large cylinders.
The existence and properties of the Rayleigh--Bloch waves are connected with finite-array resonances.
\end{abstract}


\section{Introduction}\label{sec:intro}

\begin{figure}[b]
\centering
\setlength{\figurewidth}{0.9\textwidth} 
\setlength{\figureheight}{0.8\textwidth} 

\definecolor{mycolor1}{rgb}{0.00000,1.00000,1.00000}%
\begin{tikzpicture}

\begin{axis}[%
name=myaxis,
width=\figurewidth,
height=\figureheight,
at={(0\figurewidth,0\figureheight)},
scale only axis,
axis equal,
axis equal image,
xmin=-12,
xmax=12,
xtick={},
xlabel={$x$},
ymin=-4,
ymax=4,
ytick={},
ylabel style={font=\color{white!15!black}},
ylabel={$y$},
axis background/.style={fill=blue!20!white,rounded corners},
hide axis
]


\setlength{\myxunit}{0.04755\figurewidth}
\setlength{\rad}{0.7\myxunit}
    
\filldraw[fill=lightgrey,draw=black,ultra thick] (axis cs:-5,0) circle [radius=\rad] node(C1) [align=center] {$C_{1}$}; 
\filldraw[fill=lightgrey,draw=black,ultra thick] (axis cs:-2,0) circle [radius=\rad] node(C2) [align=center] {$C_{2}$}; 
\filldraw[fill=lightgrey,draw=black,ultra thick] (axis cs:1,0) circle [radius=\rad] node(C3) [align=center] {$C_{3}$}; 
\filldraw[fill=lightgrey,draw=black,ultra thick] (axis cs:8,0) circle [radius=\rad] node(C4) [align=center] {$C_{N}$}; 

\draw [loosely dotted,ultra thick] ($(C3)!0.3!(C4)$) -- ($(C4)!0.3!(C3)$);


\draw [|-|,ultra thick] ([yshift=-1.75\rad]C2.center) -- node[below] {$1$} ([yshift=-1.75\rad]C3.center);

\draw [|-|,ultra thick,black] ([yshift=-1.75\rad]$(C1.center)-(\rad,0)$) -- node[below] {$2\,a$} ++(2\rad,0);

\coordinate (c1) at (axis cs:-11,1.5);
\coordinate (c3) at (axis cs:-11,0);
\coordinate (c2) at (axis cs:-7.5,0);

\draw [dashed,black,thick] (c3) -- ($(c3)!0.8!(c2)$);
\pic [draw, -, "\large{$\psi$}", angle eccentricity=1.25, angle radius=30pt,ultra thick,black] {angle = c1--c2--c3};
\draw [-latex,ultra thick] (c1) -- node[above,yshift=0.08\myxunit] {\Large{$\phi_{\textnormal{amb}}$}} (c2);


\draw [->,thick] ($(C4)+(2\rad,0)$) -- ++(2\rad,0) node[right,xshift=-0.1\myxunit] {$x$};
\draw [->,thick] ($(C1)+(0,2\rad)$) -- ++(0,2\rad) node[above,yshift=-0.1\myxunit] {$y$};

\end{axis}

\end{tikzpicture}%
\caption{Plan view of ambient wave and finite line array of $N$ identical cylinders.}\label{fig:schematic}
\end{figure}
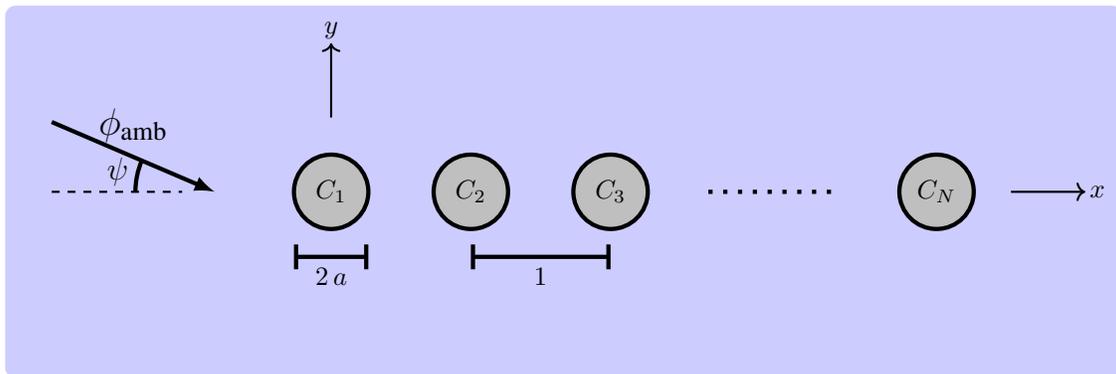

Consider an ambient plane surface gravity wave incident on an array of $N$ identical, equally-spaced vertical cylinders (Fig.~\ref{fig:schematic}), which may model, e.g., supports for offshore structures.  
Geometrical uniformity with respect to depth allows the problem to be reduced to the Helmholtz equation in the horizontal plane occupied by the water, $\phi_{xx}+\phi_{yy}+k^{2}\,\phi = 0$ for $(x,y)\in\Omega$,
where $\phi(x,y)$ is a (reduced) potential function, $k$ is a wavenumber that acts as a proxy for a prescribed angular frequency of motion (through a dispersion relation), 
plus no-normal-flow conditions on the cylinder boundaries.
For simplicity, circular cylinders of radius $a$ are assumed, which, without loss of generality, are aligned along the $x$-axis with unit centre-to-centre spacing, so that $\Omega=\{(x,y): (x-n)^2+y^{2}>a^{2}, n=1,\ldots,N\}$.

It would be reasonable to assume the wave field away from the array ends can be approximated by the corresponding infinite array ($n\in\mathbb{N}$).
But this is typically not true due to excitation of Rayleigh--Bloch waves, which propagate along the array with wavenumber $\beta(k)>k$, and decay exponentially away from it \citep{porter_rayleighbloch_1999}.
They are unforced solutions of the infinite-array problem that exist for all $k\leq{}k_{\textnormal{c}}<\pi$, where $k_{\textnormal{c}}(a)$ is known as the cut-off frequency. Plane waves cannot excite Rayleigh--Bloch waves along an infinite array as $\beta(k)>k$, but they can excite Rayleigh--Bloch waves along a semi-infinite array that propagate away from the end \citep{linton_scattering_2007,peter_water-wave_2007}, or in both directions along a finite array  \citep{thompson_new_2008}.

Rayleigh--Bloch waves are a class of Bloch wave, which are familiar in the analysis of doubly-periodic structures such as photonic/phononic crystals. 
Rayleigh--Bloch waves are the localised solutions obtained in the limit that the unit cell for a doubly-periodic structure tends to infinity in one dimension \citep{mciver_water-wave_2000}, and, thus, they are also a class of trapped mode \citep{porter_rayleighbloch_1999}.
In the limit, a continuous spectrum appears, and Rayleigh--Bloch waves are closely connected with embedded trapped modes \citep{linton_embedded_2007}, 
otherwise known as bound states in the continuum \citep{hsu_bound_2016}.
 
\begin{figure}
\centering
\setlength{\figurewidth}{0.9\textwidth} 
\setlength{\figureheight}{0.7\textwidth} 
\input{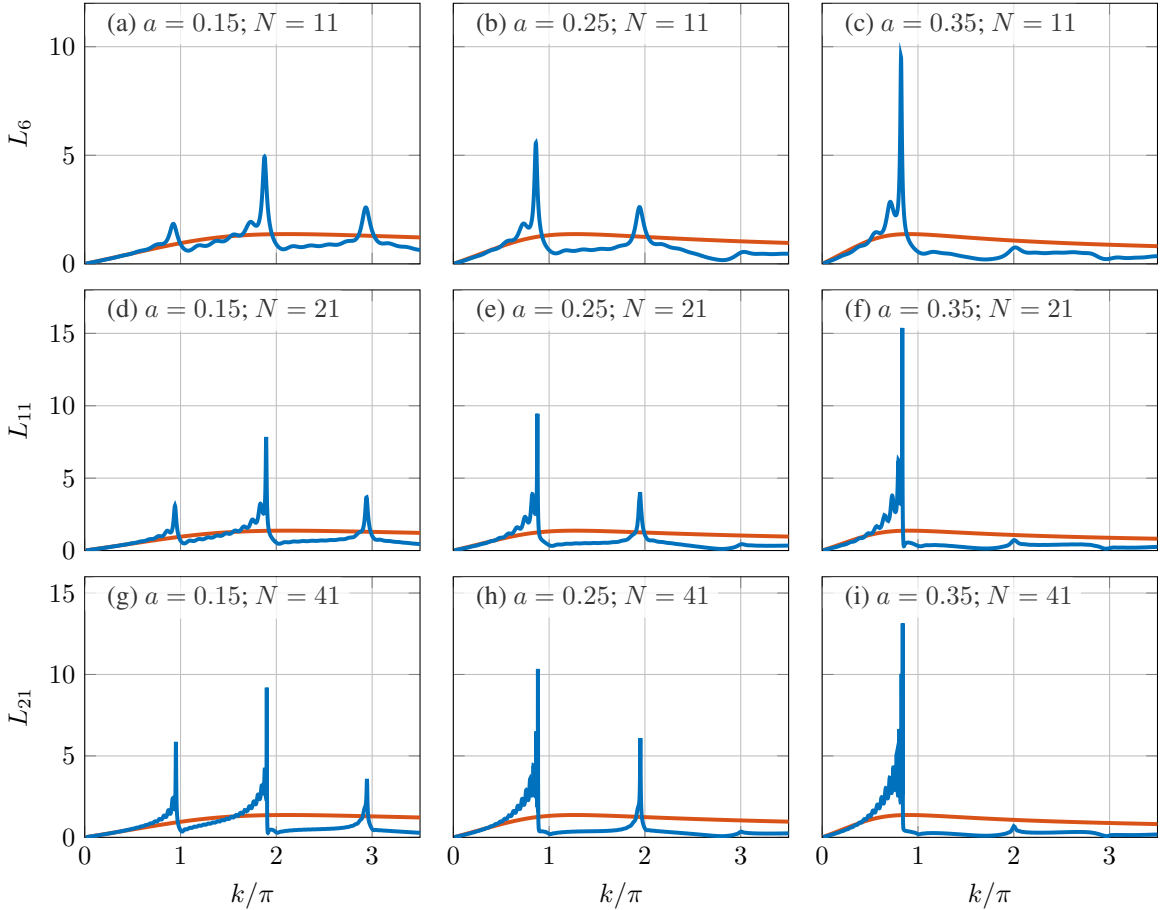}
\caption{Normalised load on middle cylinder in array vs.\ frequency (blue curves) and single cylinder reference (red), for $\psi=0$.}\label{fig:loads}
\end{figure}

\citet{maniar_wave_1997}'s discovery of resonant loads on finite arrays motivated a series of studies into Rayleigh--Bloch waves.
Fig.~\ref{fig:loads} shows the normalised $x$-component of the hydrodynamic load on the middle cylinder ($L_{n}$ for $n=\lceil{}N\,/\,2\rceil$) versus frequency, due to head-on incidence ($\psi=0$), for different values of cylinder radius and array length, and with loads on corresponding isolated cylinders for reference.
Primary resonances occur below $k=\pi$ in all cases, secondary resonances below $k=2\,\pi$ for the small and intermediate cylinder radius values, and tertiary resonances below $k=3\,\pi$ for the small-radius arrays.

\citet{maniar_wave_1997} connected primary resonances to trapped modes in channels containing a single cylinder along the centreline and Neumann conditions on the side walls, which exist at $k=k_{\textnormal{N}}(a)$ just above the resonant frequencies for all radius values. 
Similarly, they connected secondary resonances to channel modes with Dirichlet conditions applied on the channel walls, which exist at $k=k_{\textnormal{D}}(a)$ up to $a\approx{}0.3394$ \citep{evans_trapped_1998}, explaining the absence of secondary resonances for the large-radius arrays in Fig.~\ref{fig:loads}.

Rayleigh--Bloch waves are equivalent to Neumann modes in the standing wave limit $\beta\to\pi$ as $k\to{}k_{\textnormal{N}}\equiv{}k_{\textnormal{c}}$ \citep{porter_rayleighbloch_1999}.
\citet{evans_trapping_1999} gave evidence Rayleigh--Bloch waves are responsible for primary resonances, and derived the approximation $\beta\approx{}(N-1)\,\pi\,/\,N$ at the primary resonance, based on the Rayleigh--Bloch wave being a modulated Neumann mode.
The studies connecting Rayleigh--Bloch waves with primary resonances culminated in \citet{thompson_new_2008} showing resonant responses are due to constructive interference between Rayleigh--Bloch waves after reflections by the array ends.

The emergence of metamaterials is generating interest in Rayleigh--Bloch waves as a mechanism to control wave propagation.
The related structures include ring resonators \citep{maling_whispering_2016}, graded resonator arrays  \citep{bennetts_graded_2018}, flat lensing meta-arrays \citep{chaplain_flat_2019} and superhydrophobic metasurfaces \citep{schnitzer_acoustics_2019}.


Rayleigh--Bloch waves have only been found for certain parameter combinations above the cut-off \citep{porter_embedded_2005}, where they are embedded in the continuous spectrum, \ie at frequencies that permit energy radiation away from the (infinite) array. 
Thus, secondary resonances and above have not been connected with Rayleigh--Bloch waves.

In this article, we show extensions of Rayleigh--Bloch waves exist for frequencies above the cut-off, via the spectrum of a transfer operator originally proposed for random media \citep{bennetts_wave_2011,montiel_evolution_2015}.  
Existence of Rayleigh--Bloch waves and their properties are linked to resonances for the finite arrays shown in Fig.~\ref{fig:loads}.

\section{Transfer operator for a generalised channel}

Consider the cylinder $C=\{(x,y):x^{2}+y^{2}\leq{}a\}$ at the centre of a unit-width channel that occupies the horizontal domain $\Omega_{\textrm{c}}=\{(x,y):-0.5<x<0.5,y\in\mathbb{R}\}$. 
The potential function in the channel can be expressed in terms of its directional spectrum, as
\begin{equation}
\phi(x,y)=
\int_{\Gamma}a_{\pm}(\chi)\,\varphi_{\pm}(x,y:\chi)\wrt{}\chi	
+
\int_{\Gamma}b_{\pm}(\chi)\,\varphi_{\pm}(x,y:\chi+\pi)\wrt{}\chi,	
\end{equation}
for $(x,y)\in\Omega_{\pm}$,
where $\Gamma=\{\pm(\pi\,/\,2-\ci\,\gamma):\gamma\in\mathbb{R}_{+}\}\bigcup\{\gamma\in\mathbb{R}:-\pi\,/\,2<\gamma<\pi\,/\,2\}$ 
is directional space, 
$\varphi_{\pm}=\exp\{\ci\,k\,\big((x\mp{}0.5)\,\cos\chi+y\,\sin\chi\big)\}$ is a plane wave with direction $\chi$ normalised to the left ($-$) or right ($+$) of the channel, 
and $\Omega_{\pm}=\{(x,y):(x,y)\in\Omega_{\textrm{c}}, \pm{}x>0\}$.
The directional spectrum is defined by the amplitude functions $a_{\pm}$ (components propagating or decaying rightwards, \ie in the positive $x$-direction) and $b_{\pm}$ (leftwards), which 
are related via $b_{-}=\mathcal{R}\{a_{-}\}+ \mathcal{T}\{b_{+}\}$ and $a_{+}=\mathcal{T}\{a_{-}\}+ \mathcal{R}\{b_{+}\}$.
The reflection and transmission operators, $\mathcal{R}$ and $\mathcal{T}$, are, respectively,
\begin{equation}
	\mathcal{R}\{\bullet\}=\frac{1}{\pi}\int_{\Gamma}\,R(\chi:\psi)\,\bullet(\psi)\wrt\psi
	\mathand
	\mathcal{T}\{\bullet\}=\frac{1}{\pi}\int_{\Gamma}\,T(\chi:\psi)\,\bullet(\psi)\wrt\psi,
\end{equation}
\begin{subequations}
\begin{align}
	\mathXR{where}
	R(\chi:\psi)&=\e^{\frac{\ci\,k\,(\cos\chi+\cos\psi)}{2}}\sum_{m=-\infty}^{\infty}Z_{m}\,\e^{\ci\,m\,(\chi-\psi-\pi)};
	\quad
	Z_{m}=-\frac{\mathrm{J}'_{m}(k)}{\mathrm{H}'_{m}(k)}
	\\[6pt]
	\mathandR
	T(\chi:\psi)&=\e^{\frac{\ci\,k\,(\cos\chi+\cos\psi)}{2}}
	\left\{
	\delta(\chi-\psi)\,\pi
	+
	\sum_{m=-\infty}^{\infty}Z_{m}\,\e^{\ci\,m\,(\chi-\psi)}
	\right\},
\end{align}
\end{subequations}
with 
$\mathrm{J}_{m}$ and $\mathrm{H}_{m}$ order $m$ Bessel and Hankel functions of the first kind.

The transfer operator $\mathcal{P}$ for the channel is such that \citep{bennetts_localisation_2017}
\begin{equation}
	\left(
	\begin{array}{c}
	a_{+}
	\\
	b_{+}
	\end{array}
	\right)
	=
	\mathcal{P}\,
	\left(
	\begin{array}{c}
	a_{-}
	\\
	b_{-}
	\end{array}
	\right)
	\mathand
	\mathcal{P}=
	\left(
	\begin{array}{c c}
	\mathcal{T}-\mathcal{R}\,\mathcal{T}^{-1}\,\mathcal{R}
	&
	\mathcal{R}\,\mathcal{T}^{-1}
	\\
	\mathcal{T}^{-1}\,\mathcal{R}
	&
	\mathcal{T}^{-1}
	\end{array}
	\right).	
\end{equation}
Its spectrum gives infinite array solutions, 
where the eigenvalues $\mu\in\mathrm{eig}\{\mathcal{P}\}$ define the quasi-periodicities, and noting the spectrum is with respect to directional space rather than frequency space \citep[as in, e.g.,][]{porter_rayleighbloch_1999}.
The continuous spectrum defines forced solutions, \ie involving an ambient wave, and the discrete (point) spectrum defines the unforced solutions, \ie Rayleigh--Bloch waves.
The directional spectrum is truncated along the complex branches and discretised for computations \citep{montiel_evolution_2015}.

\section{Rayleigh--Bloch waves above the cut-off}\label{sec:rb-modes}

\begin{figure}
\centering
\includegraphics[width=0.8\textwidth]{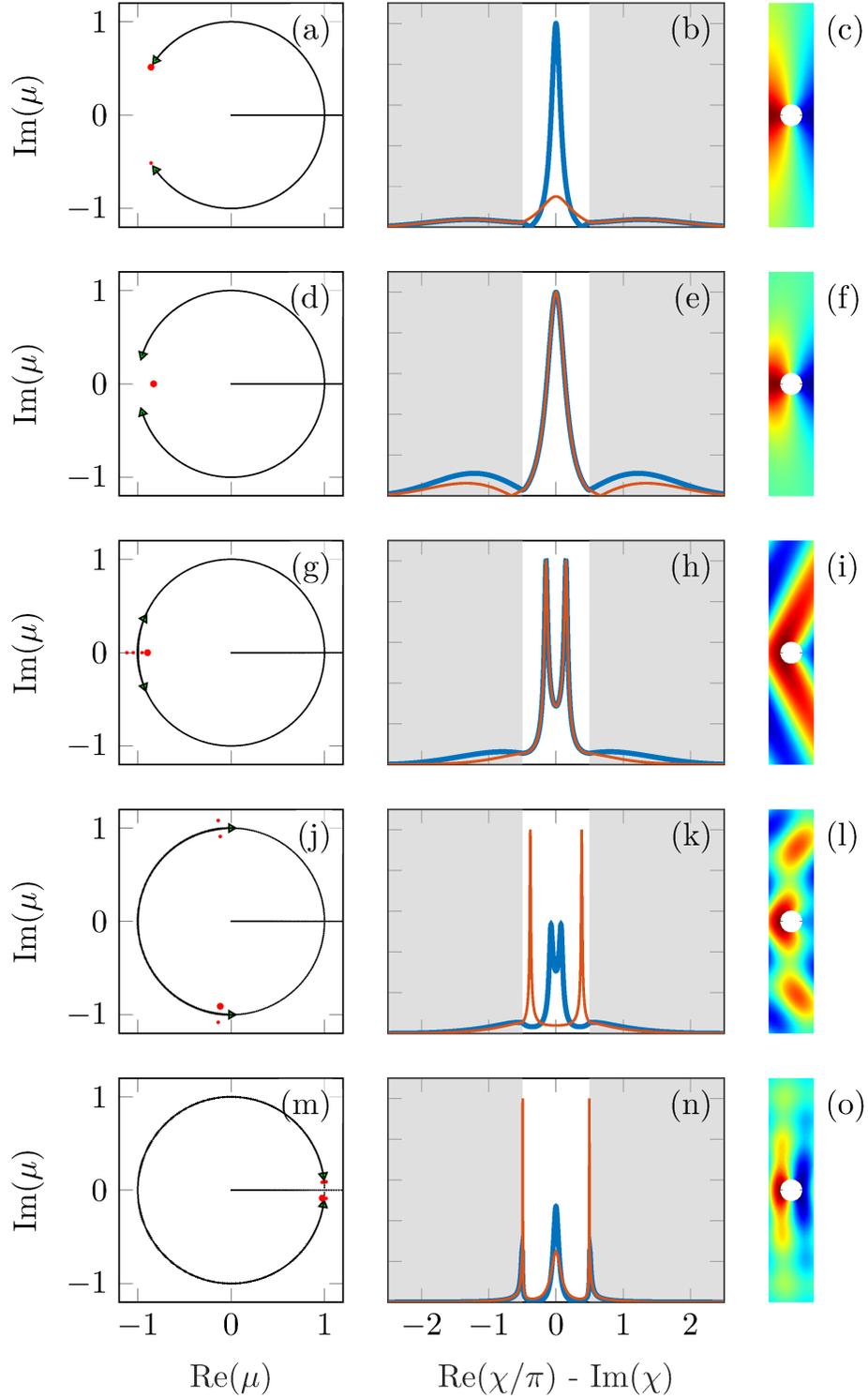}
\caption{Left-hand panels:~Continuous spectrum (small black bullets) and discrete spectrum (red bullets). 
Middle:~Directional spectrum (eigenfunctions; $a(\chi)$ thick blue and $b(\chi)$ thin red) corresponding to highlighted eigenvalue (large red bullets in left-hand panels).
Right:~Real parts of corresponding modes for $-0.5<x<0.5$ and $-5<y<5$.
Top row to bottom:~$k\,/\,\pi= 0.8$, 0.9, 1.119, 1.5, 1.95.  }\label{fig:RBmodes}
\end{figure}

Fig.~\ref{fig:RBmodes} shows spectra of $\mathcal{P}$ for radius $a=0.25$ and frequencies from $k=0.8\,\pi$ (top panels) to $1.95\,\pi$ (bottom).
The left-hand panels show eigenvalues in the complex plane, divided into an approximation of the continuous spectrum (closely spaced black bullets) and the discrete spectrum (red bullets).
The middle panels show moduli of the directional spectra (eigenfunctions) corresponding to the indicated discrete-spectra eigenvalues, and the right-hand panels show the real parts of the corresponding modes, \ie Rayleigh--Bloch waves.

Eigenvalues in the continuous spectra are $\mu\in\Upsilon_{\textnormal{c}}=\{\exp(\pm\ci\,k\,\cos\chi):\chi\in\Gamma\}$, and have multiplicity at least two due to symmetry of directional space.
Eigenvalues around the unit circle are solutions for propagating ambient waves ($\chi\in\mathbb{R}$), and those on the positive real axis for ambient waves growing/decaying in the $x$-direction ($\chi\in\mathbb{C}$).
For frequencies such that $k<\pi$, the propagating part of the continuous spectrum occupies only an arc of the unit circle with tips $\mu=\exp\{\pm\ci\,k\}$ (\ie head-on incidence, $\psi=0$), so the arc occupies a greater proportion of the unit circle as frequency increases (Fig.~\ref{fig:RBmodes}a,d).
The tips meet at $\mu=-1$ when $k=\pi$, at which point the continuous spectrum occupies all of the unit circle, and for $k>\pi$ the continuous spectrum begins to overlap itself (Fig.~\ref{fig:RBmodes}g).

For frequencies below the cut-off, $k_{\textnormal{c}}\approx{}0.89\,\pi$, the discrete spectrum contains two eigenvalues, which lie on the unit circle just ahead of the continuous-spectrum tips (Fig.~\ref{fig:RBmodes}a).
The eigenvalues are the reciprocal pair $\mu=\exp\{\pm\ci\,\beta\}$, where $\beta\in\mathbb{R}$ is the Rayleigh--Bloch wavenumber. The eigenvalues in the upper and lower halves of the complex plane define Rayleigh--Bloch waves propagating rightwards and leftwards, respectively. 
The directional spectrum for the eigenvalue in the upper-half plane has a rightward-propagating component ($a(\chi)$ for $-0.5<\chi\,/\,\pi<0.5$) that outweighs the leftward-propagating component ($b(\chi)$), which is consistent with the Rayleigh--Bloch wave propagating rightwards (Fig.~\ref{fig:RBmodes}b).
The corresponding mode is localised to the array, as anticipated (Fig.~\ref{fig:RBmodes}c).
As frequency increases, the eigenvalues in the discrete spectrum move around the unit circle ahead of the tips of the continuous spectrum. 
At the cut-off frequency, the eigenvalues meet at $\mu=-1$ ($\beta=\pi$), at which point the Rayleigh--Bloch waves cease propagating and become the Neumann mode. 

For frequencies just above the cut-off, the eigenvalues move onto the negative real axis (Fig.~\ref{fig:RBmodes}d), \ie $\beta=\pi+\ci\,\beta_{\textnormal{i}}$ where $\beta_{\textnormal{i}}\in\mathbb{R}_{+}$.
By reciprocity, one eigenvalue lies inside the unit circle and the other outside (latter is beyond axes limits).
Thus, the extensions of the Rayleigh--Bloch waves above the cut-off decay along the array, which can be inferred from the rightward- and leftward-propagating components of the directional spectra only differing along the complex branches (Fig.~\ref{fig:RBmodes}e). 
The modes remain localised to the array (Fig.~\ref{fig:RBmodes}f).
The eigenvalues move farther away from the unit circle along the negative real axis (increasing $\beta_{\textnormal{i}}$) until $k\approx{}1.01\pi$, at which point they move back towards the unit circle (decreasing $\beta_{\textnormal{i}}$).
The smooth peaks in the directional spectra around $\chi=0$ split into pairs of sharper peaks (Fig.~\ref{fig:RBmodes}h),  causing the mode to stretch farther away from the array (Fig.~\ref{fig:RBmodes}i). 
At $k\equiv{}k_{\shortrightarrow}\approx{}1.1\,\pi$, an additional reciprocal pair of eigenvalues join the discrete spectrum, emanating from the continuous spectrum at $\mu=-1$, and moving away from the unit circle along the negative real axis until they collide with the original eigenvalues at $k\equiv{}k_{\arrowscross}\approx{}1.12\,\pi$.
One eigenvalue in each collision moves into the upper-half complex plane and the other into the lower-half, so that $\beta=\beta_{\pm}=\pi\pm\beta_{\textnormal{r}}+\ci\beta_{\textnormal{i}}$ with $\beta_{\textnormal{r}},\beta_{\textnormal{i}}\in\mathbb{R}_{+}$.

The eigenvalue pairs in the upper and lower halves of the complex plane chase their respective continuous-spectrum tips around the unit circle (Fig.~\ref{fig:RBmodes}j).
For the highlighted eigenvalue, $\mu=\exp\{\ci\,\beta_{+}\}$, the peaks in the rightward propagating component of the directional spectrum tend towards $\chi=0$ and become smoother, whereas the peaks in the leftward-propagating component tend towards $\chi=\pm{}\pi\,/\,2$ and become sharper (Fig.~\ref{fig:RBmodes}k).
The mode stretches increasingly far away from the array  and complicated interference patterns develop around the array (Fig.~\ref{fig:RBmodes}l).
Just below the Dirichlet mode frequency, $k=k_{\textnormal{D}}(0.25)\approx{}1.9555\,\pi$, the eigenvalues overtake the tips of the discrete spectrum and tend towards the continuous spectrum on the unit circle (Fig.~\ref{fig:RBmodes}m).
Both rightward and leftward components of the directional spectra display smooth peaks around $\chi=0$, similar to below the cut-off, and sharp peaks close to $\chi=\pm{}\pi\,/\,2$ (Fig.~\ref{fig:RBmodes}n).
Cancellation of the sharp peaks causes the modes to regain array localisation (Fig.~\ref{fig:RBmodes}o).

Identifying the discrete spectrum becomes increasingly difficult as it tends towards the continuous spectrum. 
The discrete and continuous spectra are separated by modifying the cylinder boundary condition to include (artificial) damping, such that $\phi_{r}+\ci\,\varepsilon\,\phi=0$ for $r=a$ (where $r=\sqrt\{x^{2}+y^{2}\}$).
The sign of $\varepsilon\in\mathbb{R}$ determines which pair of eigenvalues move away from the unit circle.  
In contrast, the continuous spectrum remains in $\Upsilon_{\textnormal{c}}$  
as the ambient waves dictate the quasi-periodicities of the corresponding solutions.
Suitable choices of $\varepsilon$ make it possible to identify the discrete spectrum as the frequency passes through the Dirichlet mode.
At each frequency, homotopy is used to obtain the discrete spectrum for the original (undamped) problem, by decreasing $\vert\varepsilon\vert$ to zero in small steps and tracking the eigenvalue--eigenfunction pairs.

\begin{figure}
 \setlength{\figurewidth}{0.8\textwidth} 
 \setlength{\figureheight}{0.4\textwidth} 
 \input{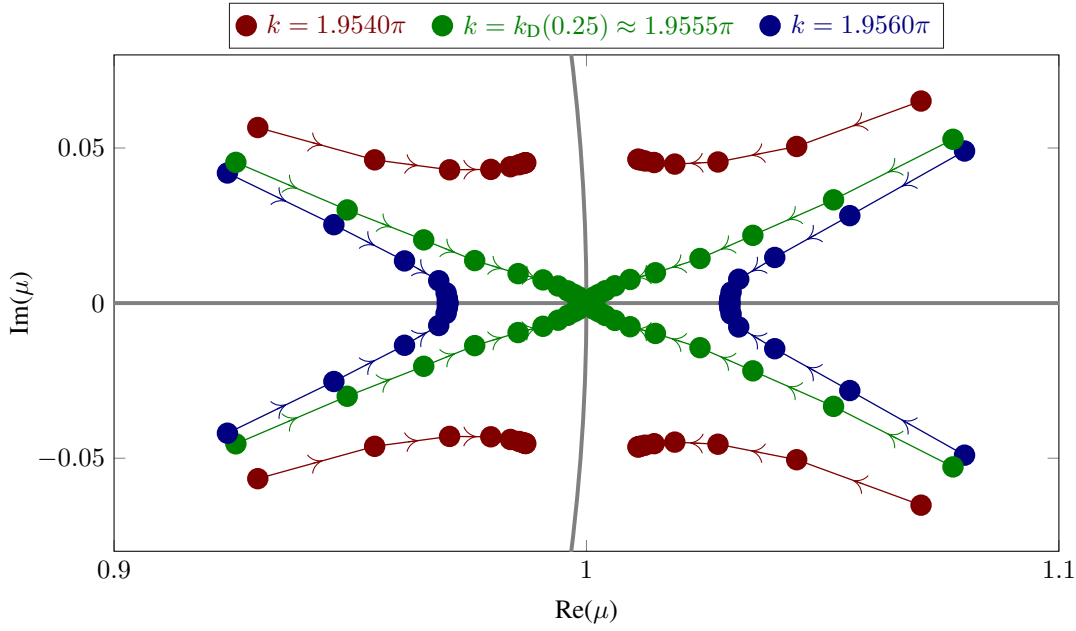}
 \caption{Discrete spectrum around Dirichlet mode frequency, for damping $\vert\varepsilon\vert=0.2$ to 0.0001 with logarithmically scaled steps, where arrows indicate decreasing $\vert{}\varepsilon\vert$.}\label{fig:homotopy} 
\end{figure}

Fig.~\ref{fig:homotopy} illustrates the homotopy method and shows the discrete spectrum below, at and above the Dirichlet mode frequency.
Below the Dirichlet mode frequency, the discrete spectrum is of the form $\mu=\exp\{\pm\ci\,\beta_{\pm}\}$ and close to the unit circle ($\beta_{\textnormal{i}}\ll{}1$).
At $k=k_{\textnormal{D}}\approx{}1.9555\,\pi$, the four eigenvalues coalesce at $\mu=1$, \ie $\beta=0$ or $2\,\pi$, at which point the extended Rayleigh--Bloch waves become the Dirichlet mode.
Above the Dirichlet mode, the discrete spectrum contains a reciprocal pair of eigenvalues, $\mu=\exp\{\mp\beta_{\textnormal{i}}\}$, which lie on the positive branch of the real axis, embedded in the continuous spectrum.
The discrete spectrum vanishes into the continuous spectrum for $k\approx{}2\,\pi$, \ie all eigenvalues $\mu\in\Upsilon_{\textnormal{c}}$ irrespective of the $\varepsilon$-value. 

\begin{figure}
 \centering
 \setlength{\figurewidth}{0.9\textwidth} 
 \setlength{\figureheight}{0.6\textwidth} 
 \input{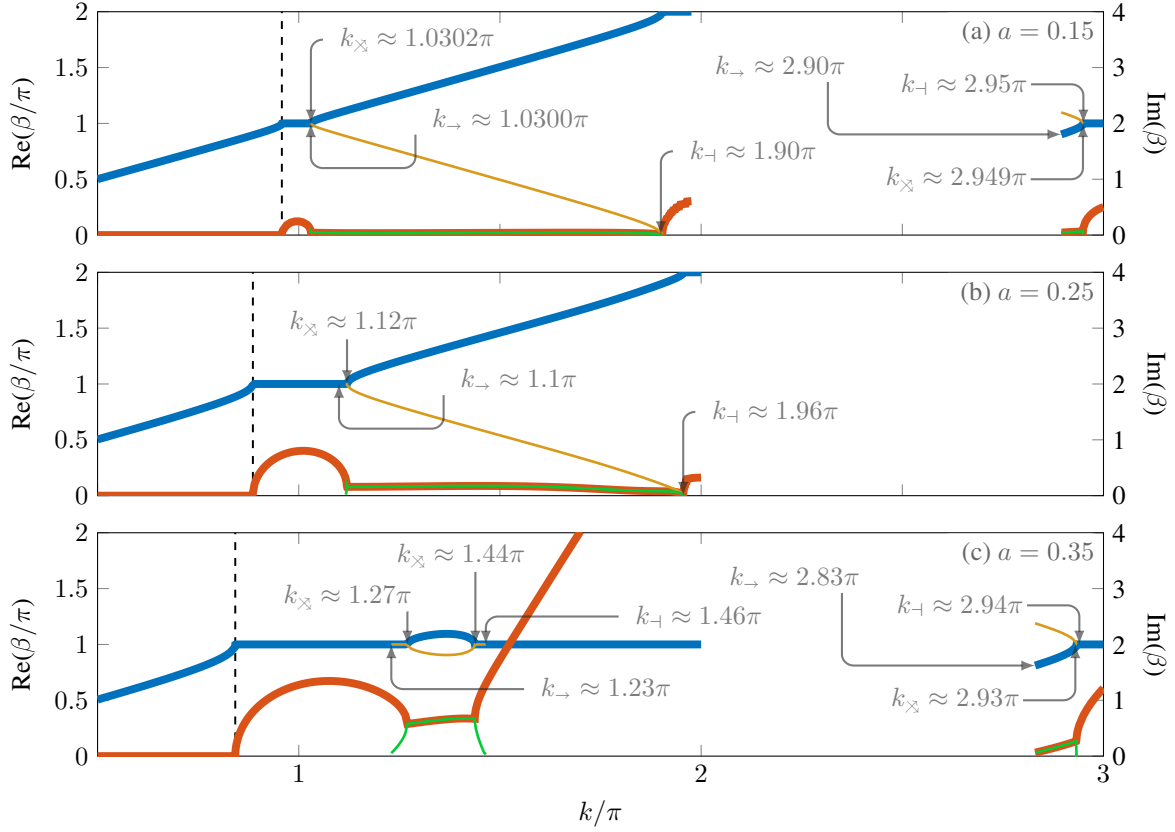}
 \caption{Rayleigh--Bloch wave dispersion curves. Real parts (thick blue and thin yellow curves) and imaginary parts (thick red and thin green).}\label{fig:dispersion}
\end{figure}

Fig.~\ref{fig:dispersion} shows the Rayleigh--Bloch dispersion curves for the small, intermediate and large radii up to $k=3\,\pi$.
The dispersion curves for the intermediate radius ($a=0.25$; Fig.~\ref{fig:dispersion}b) follow the above analysis, with a single $\beta\in\mathbb{R}$ for $k\leq{}k_{\textnormal{c}}\approx{}0.89\,\pi$, $\beta=\pi+\ci\,\beta_{\textnormal{i},j}$ ($j=1,2$) for $k_{\textnormal{c}}<k<1.12\,\pi$ where the second wavenumber appears for $k>1.1\,\pi$, $\beta=\pi\pm\beta_{\textnormal{r}}+\ci\,\beta_{\textnormal{i}}$ for $1.12\,\pi<k<1.96\,\pi$, and a single $\beta=2\,\pi+\ci\,\beta_{\textnormal{i}}$ for $1.96\,\pi<k<2\,\pi$.
There is only evidence of a continuous spectrum for $k>2\,\pi$.

The dispersion curves for the small radius ($a=0.15$; Fig.~\ref{fig:dispersion}a) are qualitatively similar to the intermediate radius for $k<2\,\pi$, with the most notable difference being the shorter interval for which $\beta=\pi+\ci\,\beta_{\textnormal{i}}$ above the cut-off.
In contrast to the intermediate radius, the discrete spectrum reappears at $k_{\rightarrow}\approx{}2.90\,\pi$, with four eigenvalues emanating from the continuous spectrum on the unit circle just behind the tips of the continuous spectrum, defined by a pair of complex-valued Rayleigh--Bloch wavenumbers $\beta=\pm\beta_{\textnormal{r}}+\ci\beta_{\textnormal{i}}$.
The eigenvalues move slightly away from the unit circle (increasing $\beta_{\textnormal{i}}$) and follow the continuous-spectrum tips around the unit circle ($\beta_{r}\to{}\pi$) until $k_{\arrowscross}\approx{}2.949\,\pi$, when pairs of eigenvalues meet on the negative real axis (one pair inside and one outside) and then move in opposite directions along the real axis, so that $\beta=\beta_{j}=\pi+\ci\,\beta_{\textnormal{i},j}$ $(j=1,2)$.
At $k_{\dashv}\approx{}2.95\,\pi$, the eigenvalues moving towards the unit circle vanish into the  continuous spectrum at $\mu=-1$, leaving only a reciprocal pair of eigenvalues in the discrete spectrum, which tend towards the origin from below and negative infinity (increasing $\beta_{\textnormal{i}}$).

The dispersion curves for the large radius ($a=0.35$; Fig.~\ref{fig:dispersion}c)  are qualitatively similar to those for the smaller radii for frequencies only up to where the pair of eigenvalues that emerge from the continuous spectrum above the cut-off collide with the existing eigenvalues at $k_{\arrowscross}\approx{}1.27\,\pi$ and move into the complex plane.
Rather than following the tips of the continuous spectrum around the unit circle, the eigenvalues move back towards the negative real axis where they collide for a second time and then move in opposite directions along the negative real axis ($\beta=\pi+\ci\,\beta_{\textnormal{i},j}$).
The eigenvalues moving towards the unit circle vanish into the continuous spectrum at $k_{\dashv}\approx{}1.46\,\pi$ ($\beta_{\textnormal{i}}\to{}0$), whereas the remaining eigenvalue pair tend towards the origin/infinity (increasing $\beta_{\textnormal{i}}$).
The discrete spectrum reappears at $k_{\rightarrow}\approx{}2.83\,\pi$ and has similar qualitative behaviour to the discrete spectrum for $a=0.15$ below $k=3\,\pi$.

\section{Connection to resonances along finite arrays}\label{sec:resonances}

\begin{figure}
 \setlength{\figurewidth}{0.9\textwidth} 
 \setlength{\figureheight}{0.7\textwidth} 
 \input{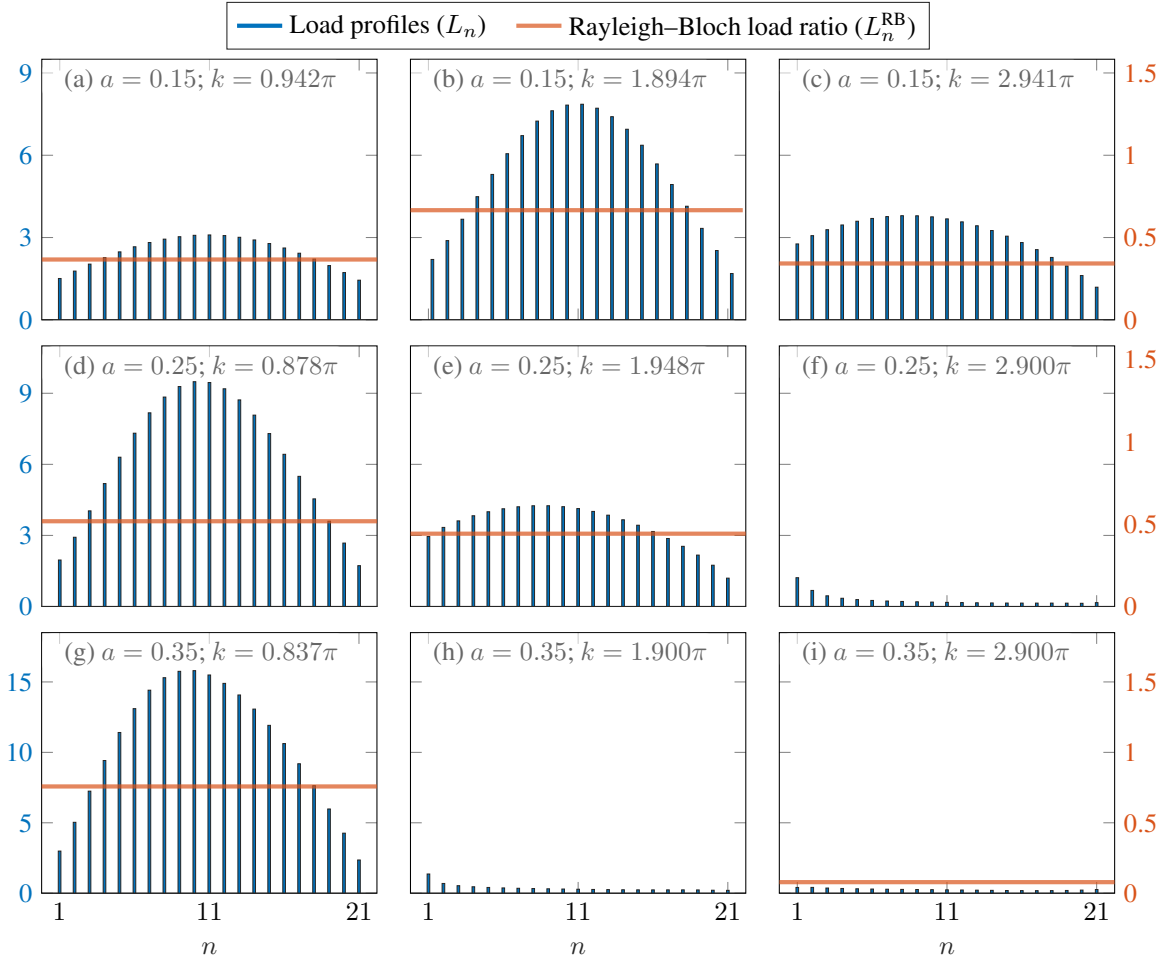}
 \caption{Load profiles along 21-cylinder arrays for $\psi=0$ (blue bars). Ratio of loads imposed by leftwards propagating/decaying Rayleigh--Bloch waves to rightwards propagating/decaying Rayleigh--Bloch waves (red lines), where, above the cut-off, dominant Rayleigh--Bloch waves are used and the ratios are scaled by $\exp\{\pm2\,\beta_{\textnormal{i}}\,(n-1)\}$.}\label{fig:resons}
\end{figure}

Fig.~\ref{fig:resons} shows load profiles along $21$-cylinder arrays for $a=0.15$, 0.25 and 0.35, at frequencies below $k=\pi$, $2\,\pi$ and $3\,\pi$.
The frequencies below the cut-offs (left-hand panels) are at the primary resonance peaks (cf.~Fig.~\ref{fig:loads}d--f). 
In each case, the rightward-propagating Rayleigh--Bloch wave is more strongly generated than the leftward-propagating Rayleigh--Bloch wave, but they are of comparable magnitude, as indicated by the ratio of the loads they impose on the cylinders ($L_{n}^{\textnormal{RB}}$).
Coherence between the rightward- and leftward-propagating Rayleigh--Bloch waves near the array centre ($\beta\approx{}(N-1)\,\pi\,/\,N$) creates the characteristic near symmetric resonant load profiles \citep{thompson_new_2008}.

For the small and intermediate radius cylinders, the frequencies below $k=2\,\pi$ (Fig.~\ref{fig:resons}b,e) are at the secondary resonance peaks. 
Four extended Rayleigh--Bloch waves exist, decaying rightwards (wavenumbers $\pi\pm\beta_{\textnormal{r}}+\ci\,\beta_{\textnormal{i}}$) or leftwards ($\pi\pm\beta_{\textnormal{r}}-\ci\,\beta_{\textnormal{i}}$) along the array, and the resonant frequencies coincide with coherence, such that $\beta_{\textnormal{r}}\approx{}(N-1)\,\pi\,/\,N$. 
The dominant Rayleigh--Bloch waves decaying rightwards and leftwards have wavenumbers $\pi+\beta_{\textnormal{i}}\pm\ci\,\beta_{\textnormal{i}}$. They are of comparable magnitude, as indicated by the $L_{n}^{\textnormal{RB}}$-values (noting the scaling of the ratios makes them constant along the array).   
For the small radius the resonant load profile is similar to those below the cut-off, but for the intermediate radius the profile is skewed towards the array front due to skewing of the coherence (not shown).

For the large radius, no resonance or extended Rayleigh--Bloch waves exist below $k=2\,\pi$ (Fig.~\ref{fig:resons}h). 
The chosen frequency is close to those for $a=0.15$ and 0.25, but the cylinder loads are small and the profile tends to decrease along the array, as diffraction dominates the response in the absence of Rayleigh--Bloch waves.

Only the small radius array has a resonance below $k=3\,\pi$, and the load profile at the resonance peak is shown (Fig.~\ref{fig:resons}c).
Four Rayleigh--Bloch waves exist, with wavenumbers $\pm(\pm\beta_{\textnormal{r}}+\ci\,\beta_{\textnormal{i}})$, such that $\beta_{\textnormal{r}}\approx{}(N-1)\,\pi\,/\,N$. 
The dominant Rayleigh--Bloch waves have wavenumbers $\beta_{\textnormal{r}}\pm\ci\,\beta_{\textnormal{i}}$, and, as with the previous resonances, they have comparable magnitudes.
The load profile is skewed towards the array front, similar to the intermediate radius resonance below $k=2\,\pi$ (Fig.~\ref{fig:resons}e).

Representative load profiles are shown for the intermediate- and large-radius arrays below $k=3\,\pi$. 
Both responses appear to be diffraction dominated, similar to the large-radius array  below $k=2\,\pi$.
For the large radius, extended Rayleigh--Bloch waves exist below $3\,\pi$, and the frequency is chosen to satisfy the coherence condition. 
However, the ratio of loads imposed by the dominant  Rayleigh--Bloch waves shows the magnitudes of the leftward-decaying waves are an order of magnitude less than the rightward-decaying waves ($L_{n}^{\textnormal{RB}}=0.079$), which precludes generation of resonance.

\section{Conclusions} 

Extensions of Rayleigh--Bloch waves above the cut-off have been calculated up to $k=3\,\pi$.
Two additional Rayleigh--Bloch waves were found to emanate from the continuous spectrum and form a symmetric quartet of complex-valued wavenumbers.
For small to intermediate radius cylinders, the extensions were shown to form continuous connections between the Neumann mode (at the cut-off frequency) and the Dirichlet mode, where the four Rayleigh--Bloch wavenumbers coalesce.
Between the Neumann and Dirichlet modes, the Rayleigh--Bloch waves were found to develop spikes in their directional spectra and lose array localisation.
Above the Dirichlet mode, the Rayleigh--Bloch waves become embedded in the branch of the continuous spectrum associated with decaying ambient waves before they vanish.
For a small radius cylinder, they were found to reappear below $k=3\,\pi$, and similar behaviour was found for a large radius cylinder. 
Existence of Rayleigh--Bloch waves was connected with resonances on finite arrays occurring for some radius values and not others, although an example was given to show that existence of Rayleigh--Bloch waves with wavenumbers satisfying a coherence condition does not guarantee resonance. 

The discovery of Rayleigh--Bloch waves above the cut-off creates the possibility of designing metamaterial-esque line arrays to control high frequency waves.
Moreover, the approach used in this study could be combined with the method of \citet{bennetts_low-frequency_2019} to  seek extensions of Rayleigh--Bloch-type waves at high frequencies for multiple-line arrays. 
\\ 

\noindent{}\textbf{Funding.} The work is supported by the Alexander von Humboldt Foundation.
LGB is supported by the Australian Research Council (FT190100404).

\bibliographystyle{unsrtnat}
\bibliography{MyBibli.bib}  






\end{document}